# Large language models accurately predict public perceptions of support for climate action worldwide

Nattavudh Powdthavee

*Nanyang Technological University*

Sandra J. Geiger

*Princeton University*

**Abstract**

Although most people support climate action, widespread underestimation of others' support stalls individual and systemic changes. In this preregistered experiment, we test whether large language models (LLMs) can reliably predict these perception gaps worldwide. Using country-level indicators and public opinion data from 125 countries, we benchmark four state-of-the-art LLMs against Gallup World Poll 2021/22 data and statistical regressions. LLMs, particularly Claude, accurately capture public perceptions of others' willingness to contribute financially to climate action (MAE ≈ 5 p.p.; r = .77), comparable to statistical models, though performance declines in less digitally connected, lower-GDP countries. Controlled tests show that LLMs capture the key psychological process—social projection with a systematic downward bias—and rely on structured reasoning rather than memorized values. Overall, LLMs provide a rapid tool for assessing perception gaps in climate action, serving as an alternative to costly surveys in resource-rich countries and as a complement in underrepresented populations.

**Main**

Addressing climate change requires rapid, coordinated changes in individual behavior and in broader social and political systems[1,2]. Recent opinion polls show that the public supports such changes. Worldwide, 69% are willing to contribute 1% of their income to help address climate change, and 89% call for more ambitious government action[3]. Yet despite broad support, current mitigation efforts remain insufficient[4,5].

Multidisciplinary research on cooperation suggests that progress on climate action depends less on personal preferences than on *beliefs* about others' support and actions[6–12]. However, these second-order beliefs are often inaccurate (pluralistic ignorance[13–15]). People consistently underestimate how many others agree that anthropogenic climate change is real[16,17], are concerned about it[18–21], support transformative climate policies[22–27], and engage in mitigation behaviors[28–30]. For example, even though a global supermajority is willing to contribute a share of their income to climate action, people believe that only 43% are willing to do so[3].

Such underestimations are a key barrier to climate action. Because individuals are more likely to contribute to collective action when they believe others are doing so (conditional cooperation)[31–33], misperceptions can suppress expressed support[17,19,34] and action, reinforcing a self-perpetuating spiral of silence and inaction[35–37]. Correcting misperceptions can help reverse this dynamic and stimulate climate discussion and engagement[3,36]. Although the evidence is mixed[10,17,19,27,38–41], interventions tend to be most effective where perception gaps are large[42–44]. Designing likely effective interventions, therefore, requires timely global maps of perception gaps. Yet large-scale measurement remains costly and logistically challenging, as it depends on paired measures of personal and perceived attitudes from high-quality, representative samples.

Large language models (LLMs) offer a potential scalable alternative or complement to large-scale surveys for rapid global assessment. Although their ability to predict human judgments varies across domains and populations[45–51], LLMs can capture patterns of public opinion about climate change[52] (when conditioned on demographics and relevant covariates) and mitigation policies[53], especially in high Human Development Index (HDI) countries, where richer training data make public opinion easier to predict[53]. However, forecasting perception gaps poses a uniquely demanding test, as it requires both first-order reasoning ('*I think that you think*') and second-order reasoning ('*I think that you think that others think*'[54]). While LLMs trained on socially rich corpora can exhibit second-order reasoning in controlled settings[55–58], they typically perform better at first-order tasks[59,60].

To date, LLM-based climate forecasts have focused exclusively on first-order reasoning, leaving it unclear whether their predictive ability extends to culturally variable second-order judgments. Here, we test whether LLMs can map global perceptions of others' climate action and identify where their predictions are most accurate. In a [preregistered](preregistered) experiment, we prompted four leading LLMs—GPT-4o mini, Claude 3.5 Haiku, Gemini 2.5 Flash, and Llama 4 Maverick (17B)—to estimate public perceptions of others' willingness to contribute financially to address climate change on a 0% (no one) to 100% (everyone) scale. Models generated predictions under eight input conditions (stages 1–8), systematically varying country-level sociodemographic, macroeconomic, and climate-related information. We benchmarked their performance against (a) 'ground truth' perception gaps observed in 125 countries in the Global Climate Change Survey[3] ($n$ = 129,902 respondents; $Md$ = 1,000 per country) from the Gallup World Poll 2021/22, and (b) ordinary least squares (OLS) and Lasso models, using out-of-sample evaluation. We further conducted non-preregistered experiments to probe how LLMs generate predictions, assessed generalizability across prompt formats, and tested performance across diverse U.S. climate policies. Overall, these analyses provide a rigorous test of whether—and where—LLMs can model complex perceptions of others' climate action and help identify where misperception-correction efforts are most needed and likely to be effective.

**Results**

**Predicting perception gaps across 125 countries using LLMs**

We first map perception gaps about climate action using the Global Climate Change Survey[3] (Fig. 1A) and compare them with LLM-based estimates (Fig. 1B). Consistent with prior work[3], perception gaps are widespread: respondents underestimate others' willingness to contribute 1% of their household income to address climate change by an average of 28.70 percentage points (p.p.; 95% CI: 27.17−30.24) on average. Underestimations are largest in parts of Latin America (e.g., Peru: −36.69 p.p., Argentina: −24.10 p.p.), West Africa (e.g., Guinea: −44.29 p.p., Nigeria: −27.90 p.p.), South Asia (e.g., Afghanistan: −41.50 p.p., Pakistan: −16.70 p.p.), and Southeast Asia (e.g., Indonesia: −36.50 p.p., Thailand: −12.00 p.p.), while several Western European countries (e.g., Germany: −28.40 p.p., France: −12.59 p.p.) exhibit smaller gaps (Fig. 1A).

All LLMs recover the broad spatial structure of these perception gaps (Fig. 1B) when provided with the full contextual information (stage 8), including sociodemographic, macroeconomic, and climate-related information. However, predicted global and national levels differ across models (Figs. B and C). Claude most closely matches the observed global mean (27.26 p.p., 95% CI: 25.51−29.00), with country-level estimates strongly aligned with survey gaps ($r = 0.77$, 95% CI: 0.69−0.84, $p < .001$; Fig. 1C). Llama performs similarly (24.03 p.p., 95% CI: 22.57−25.49; $r = 0.67$, 95% CI: 0.55–0.75, $p < .001$), whereas accuracy is lower for GPT (15.44 p.p., 95% CI: 14.41−16.47; $r = 0.60$, 95% CI: 0.47–0.69, $p < .001$) and especially Gemini (14.20 p.p., 95% CI: 13.40−15.00; $r = 0.15$, 95% CI: -0.03–0.31, $p = .103$).

Mean absolute errors (MAEs)—the absolute difference between observed and predicted perceptions of others' climate action—mirror these patterns (Fig. 1D). Using an 80:20 train–test split over 10 iterations, Claude is most accurate (mean test MAE = 4.60 p.p., 95% CI: 3.72–5.48), performing comparably to OLS (4.79 p.p., 95% CI: 4.23–5.26) and Lasso models (4.66 p.p., 95% CI: 3.72–5.48; Fig. 1D(a)). Llama shows similar performance (MAE = 7.00 p.p., 95% CI: 5.93–8.06), whereas GPT and Gemini substantially smaller than they actually are in most countries (Fig. 1C), exhibiting comparably large prediction errors (GPT: MAE = 13.23 p.p., 95% CI: 11.86–14.61; Gemini: MAE = 14.77 p.p., 95% CI: 13.64–15.89). These conclusions also hold for an alternative accuracy metric—root-mean-square errors (RMSEs; Fig. 1D(b)).

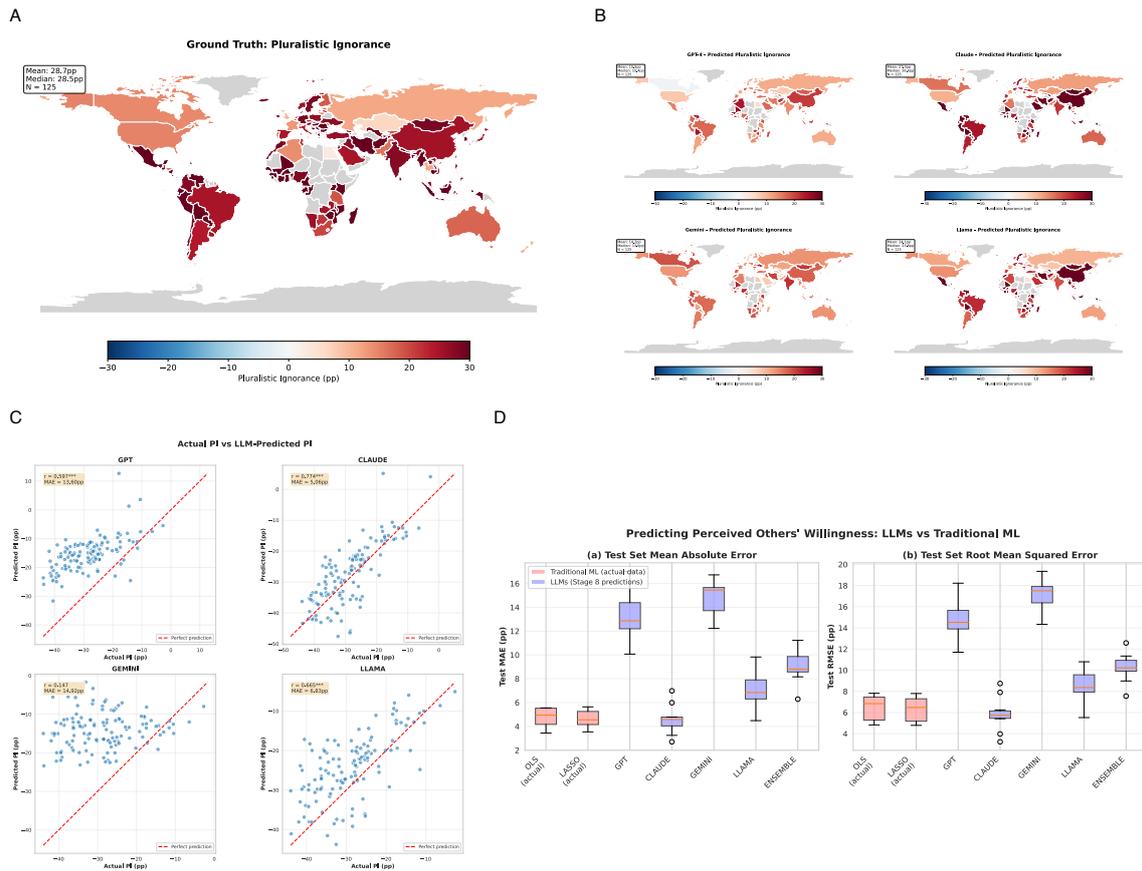

**Fig. 1. Global perception gaps and model performance.** Panel (*A*) – Ground-truth unweighted country-level perception gaps (personal willingness to contribute financially to address climate change – perceptions of others' willingness), based on 125 countries in the Global Climate Change Survey as part of the Gallup World Poll 2021/22. Panel (*B*) – LLM-estimated perception gaps using country names, average sociodemographic, macroeconomic, and climate-related variables, including mean personal willingness to contribute financially to address climate change, as input (stage 8). Panel (*C*) – Country-level scatterplots comparing ground-truth to LLM-predicted perception gaps. Panel (*D*) – Performance benchmarking for predicting perceptions of others' willingness to contribute financially to address climate change across models: (*a*) test-set mean absolute error (MAE) and (*b*) test-set root mean square error (RMSE). Traditional machine learning models (OLS and Lasso) trained on $n = 119$ countries with complete sociodemographic, macroeconomic, climate-related data, including personal willingness; LLM predictions use all available countries ($N = 125$) per panel.

## Where do LLMs predict most accurately?

We next examine whether prediction accuracy varies across countries based on structural indicators of information availability (Fig. 2). Based on prior research[45,51,61,62], LLMs are expected to make more accurate predictions in resource-rich, digitally connected countries because they are primarily trained on data from these countries.

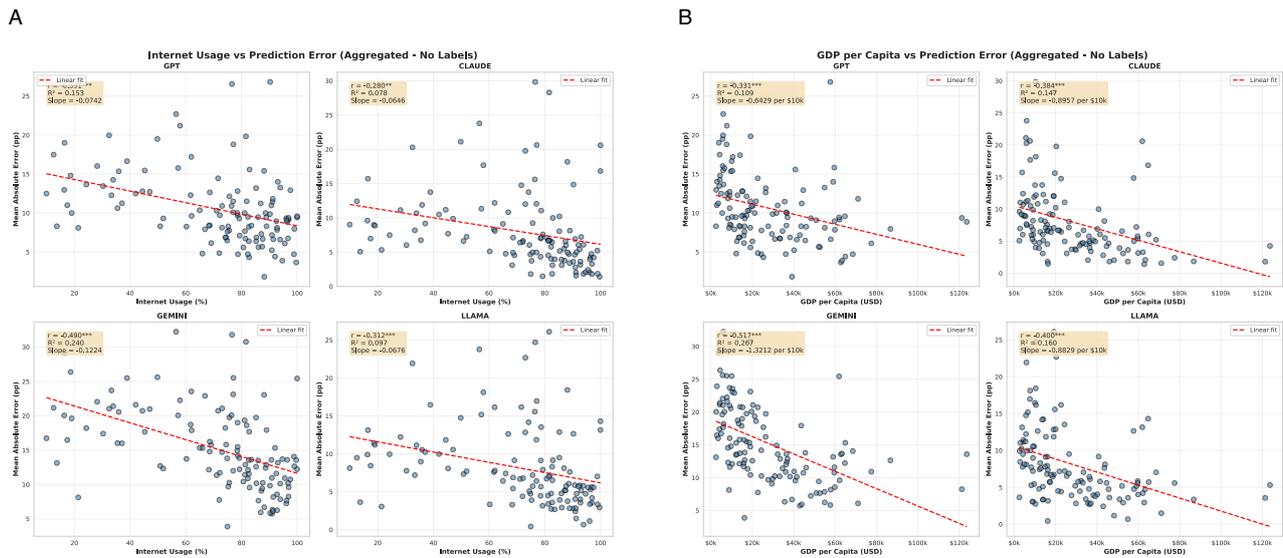

**Fig. 2. Structural predictors of model uncertainty.** Panels show the association between LLM prediction error (MAE) and two macro-structural features: (*A*) national internet penetration (% of population) and (*B*) GDP per capita (PPP, 2021). Each point represents a country; red lines reflect linear fits.

Panels A–B confirm this. LLM predictions become more accurate with both higher internet penetration and GDP per capita (all $p < .001$; $R^2$ ranges from 0.08 to 0.24 for internet penetration and from 0.11 to 0.27 for GDP per capita), with the steepest improvement observed for Gemini ($r = -0.49$ and $-0.52$ in the internet usage and GDP per capita regressions, respectively). More broadly speaking, LLMs perform best in resource-rich, digitally connected countries and provide less accurate predictions in poorer or less connected settings, reflecting representation bias in global training data.

**How do LLMs infer public perceptions of others' climate action?**

Having shown that Claude and Llama closely approximate global perception gaps about climate action, we examine how these predictions are generated. Accuracy alone cannot distinguish genuine inference from reliance on pre-training data, particularly given post-release training of some models, and system prompts do not rule out retrieval-like behavior (see Methods; Supplementary S1).

To distinguish genuine inference from memorization, we conducted four tests: (1) compare observed and LLM-implied country-level predictors (Table 1); (2) vary input information across eight stages (stages 1–8), from country alone to full information (preregistered; Fig. 3); (3) perform feature ablation to isolate each variable's contribution (Fig. 4A); and (4) run counterfactual tests, swapping data between countries to examine whether LLMs rely on country names or feature values (Fig. 4B).

**Which predictors of public perceptions of others' climate action do LLMs accurately represent?**

Using OLS regressions, we find that people in countries with (a) higher mean willingness to contribute financially to climate action ($b = 0.79$, SE = 0.08, $p < .001$) and (b) higher inequality ($b = 0.13$, SE = 0.07, $p = 0.043$) report stronger perceptions of others' willingness to contribute to climate action. All other country-level predictors are non-significant (Table 1). Importantly, LLMs recover the direction of known predictors (e.g., mean age, mean education, HDI) but often underweight socioeconomic factors. The best-performing LLMs—Claude and Llama—correctly identify country-level willingness as the key predictor of perceptions of others' willingness, without exaggerating its importance. This pattern suggests reliance on conceptual understanding rather than surface linguistic associations, consistent with social projection in second-order reasoning[63].

**Table 1**: OLS regressions predicting perceptions of others' climate action: Ground truth vs. LLMs (stage 8 with all predictors)

|  | (1) Actual | (2) GPT | (3) Claude | (4) Gemini | (5) Llama |
| --- | --- | --- | --- | --- | --- |
| Mean personal willingness to contribute financially to address climate change | 0.788*** (0.081) | 0.961*** (0.061) | 0.787*** (0.094) | 0.954*** (0.033) | 0.836*** (0.057) |
| Mean age | -0.095 (0.114) | -0.088 (0.059) | -0.108 (0.078) | -0.039 (0.038) | -0.031 (0.065) |
| Mean education level | 0.065 (0.090) | 0.154** (0.066) | 0.022 (0.094) | 0.100** (0.045) | 0.018 (0.092) |
| Mean religiosity | -0.111 (0.087) | 0.044 (0.061) | 0.104 (0.085) | 0.033 (0.040) | 0.077 (0.086) |
| HDI index (2021) | 0.123 (0.084) | 0.031 (0.061) | 0.178 (0.112) | -0.040 (0.060) | 0.007 (0.125) |
| GDP per capita (2021) | -0.089 (0.080) | 0.106 (0.072) | 0.138 (0.087) | 0.049 (0.042) | 0.204** (0.097) |
| Share of income held by the top 1% (2021) | 0.133** (0.065) | -0.013 (0.044) | 0.030 (0.072) | -0.047 (0.033) | -0.065 (0.067) |
| Average temperature 2010–2019 | -0.078 (0.097) | -0.004 (0.059) | 0.071 (0.103) | 0.002 (0.040) | 0.024 (0.073) |
| Constant | 0.000 | -0.000 | -0.000 | -0.000 | -0.000 |
| | (0.059) | (0.038) | (0.059) | (0.031) | (0.056) |
| Observations | 125 | 125 | 125 | 125 | 125 |
| $R^2$ | 0.591 | 0.827 | 0.587 | 0.884 | 0.635 |
| Adjusted $R^2$ | 0.563 | 0.815 | 0.559 | 0.876 | 0.610 |

*Note*. Standard errors in parentheses. $^{**} p < .05$, $^{***} p < .01$

**Which information do LLMs require to accurately predict public perceptions of others' climate action?**

*Sequential information addition.* Results above used the full model (stage 8), which includes country-level sociodemographic (age, education, religiosity), macroeconomic (GDP, inequality, HDI), and climate-related (temperature, personal willingness) information. To identify which inputs improve LLM predictions of perceived willingness to act on climate change, we add information sequentially (Table 1). With only country names (stage 1), most LLMs show relatively low prediction errors (Claude: 7.56 p.p., 95% CI: 6.38–8.83; Llama: 8.12 p.p., 95% CI: 7.07–9.28; GPT: 8.99 p.p., 95% CI: 7.88–10.13), except Gemini (13.0 p.p., 95% CI: 11.49–14.60) (Fig. 2). For the best-performing models, error remains stable when less relevant information is added (stages 2–4 and 6–8), but increases for worse-performing LLMs, suggesting sensitivity to irrelevant inputs. The top-performing LLMs achieve the highest accuracy (Claude: 5.35 p.p., 95% CI: 4.68–6.07; Llama: 5.35 p.p., 95% CI: 4.57–6.19) when first-order information is added alongside country names (stage 5). As GPT (13.82 p.p., 95% CI: 12.56–15.13) and Gemini (14.89 p.p., 95% CI: 13.36–16.46) overweight this information (Table 1), their predictions become less accurate. Overall, country names alone yield reasonably accurate predictions for Claude and Llama, which improve further when first-order information is provided.

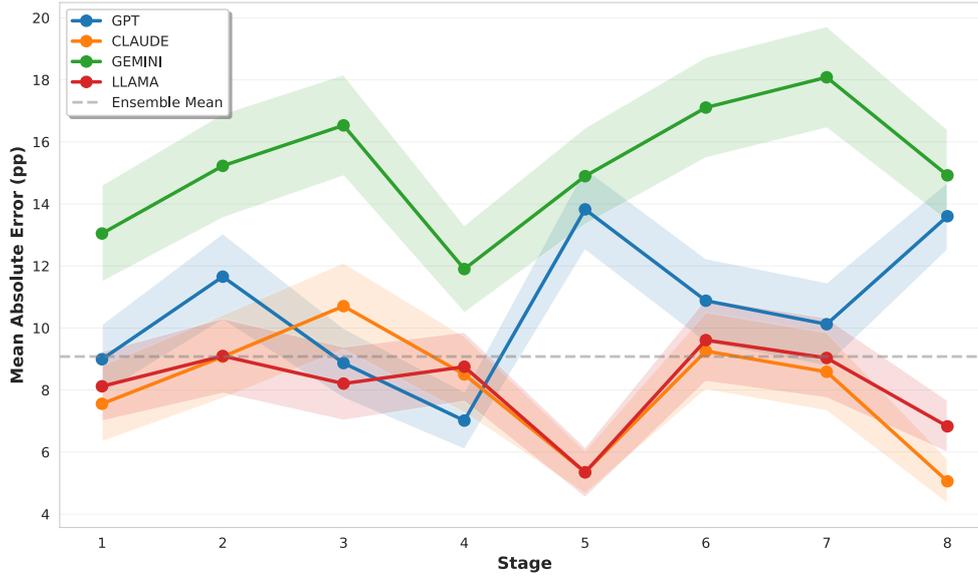

**Fig.3. Mean absolute error by information stage.** The figure shows mean absolute error (MAE) in percentage points for each LLM's predictions of others' willingness to act on climate change across eight information stages. Stage 1 includes only the country names; Stage 2 adds socio-demographic information; Stage 3 adds macro-economic indicators; Stage 4 adds temperature data; Stage 5 adds first-order willingness to act; Stage 6 combines socio-demographic and macro-economic information; Stage 7 adds temperature to Stage 6; and Stage 8 includes all available information. The horizontal dashed line represents the ensemble model's mean MAE across all stages. Lower values indicate more accurate predictions. $N$ = 125 countries × 8 stages = 1,000 predictions per model. The shaded area represents 95% confidence intervals. Ensemble mean = MAE, averaged across all models and all eight stages.

*Feature ablation.* While adding first-order information—personal willingness—improves predictions, the sequential design cannot determine whether this improvement depends on existing features. Feature ablation addresses this by removing each variable individually from the full model, isolating its independent contribution. Ablation results (Fig. 4A) show that Claude (MAE = 5.07 p.p., 95% CI: 4.37–5.79 to 5.28 p.p., 95% CI: 4.61–6.02) maintains consistently low prediction errors when less important country-level predictors of perceptions of others' climate action, such as sociodemographic, macroeconomic, and climate-related indicators, are separately omitted. Llama is slightly more sensitive to feature removal, with errors ranging from 5.41 p.p. (95% CI: 4.76–6.09) to 8.20 p.p. (95% CI: 7.27–9.17) across ablations. Removing first-order willingness ("no own willingness") increases errors by 1.50 p.p. (95% CI: 0.73–2.22) for Claude and 2.20 p.p. (95% CI: 0.06–3.44) for Llama, demonstrating that both models rely on first-order willingness to infer second-order willingness. Both best-performing LLMs distinguish between more and less relevant predictors, whereas Gemini and GPT show consistently high errors across all ablations (Gemini: 14.31–18.80 p.p.; GPT: 11.86–15.85 p.p.), suggesting they cannot effectively use the provided features.

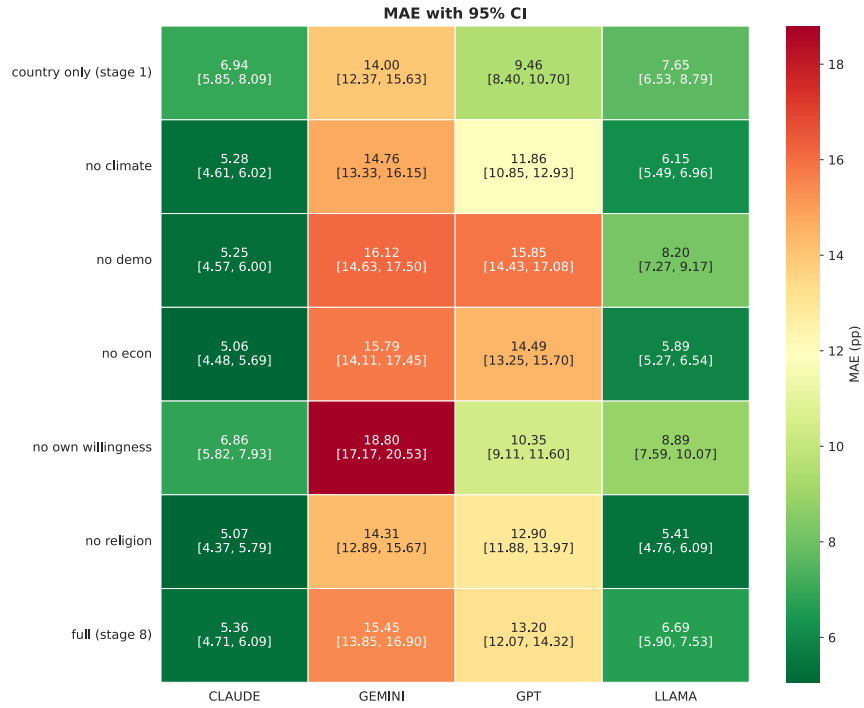

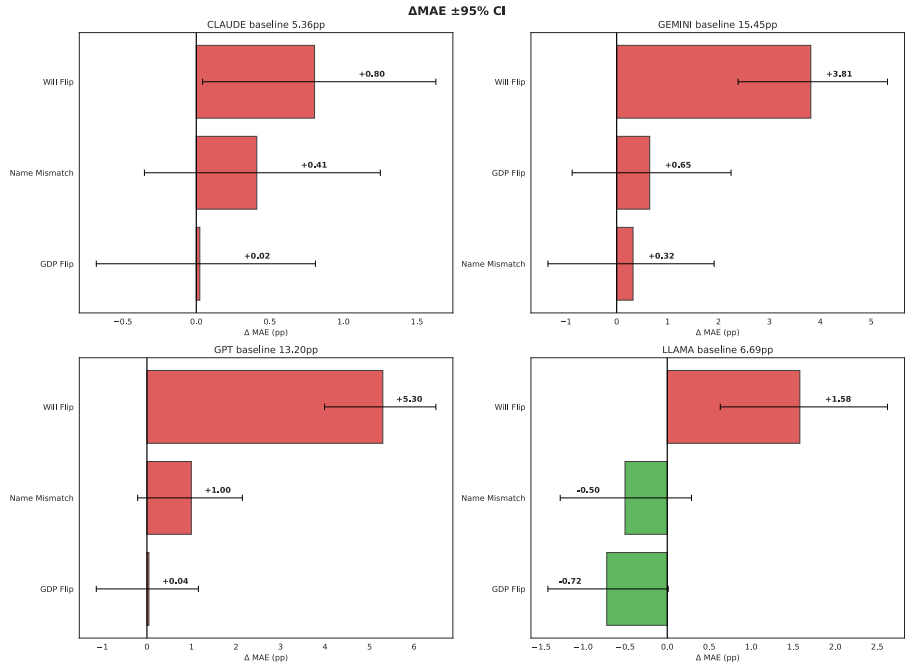

**Fig. 4. Mechanistic evaluation of LLM prediction strategies.** Panel (*A*) – Ablation tests: Heatmap showing mean absolute errors (MAEs) with 95% confidence intervals (CI) for each LLM under feature removal. Larger increases reflect greater reliance on that feature. Panel (*B*) – Counterfactual robustness tests. Own willingness–flip randomizes personal willingness across countries; name–mismatch assigns incorrect country labels while keeping other attributes constant; and GDP–flip randomizes economic values across countries. Performance stability under perturbation indicates potential reliance on memorized associations rather than structured inference.

*Counterfactual tests.* To test whether LLMs infer public perceptions of others' climate action and their predictors through structural reasoning or retrieval of stored values, we conducted three counterfactual manipulations (Fig. 4B): (1) reversing personal willingness (high ↔ low), (2) swapping country names while keeping all features unchanged, and (3) reversing GDP values (high ↔ low). If models rely on memorization, predictions should follow the country names; if they use structural reasoning, predictions should change according to the input data.

Flipping personal willingness substantially increases MAEs for Gemini (+3.81 p.p., 95% CI: -2.10–2.20, $p < .001$), GPT (+5.30 p.p., 95% CI: -1.77–1.88, $p < .001$), and Llama (+1.58 p.p., 95% CI: -1.29–1.31, $p = .018$), confirming that these models rely on first-order information to make predictions about perceptions of others' willingness. Claude shows a smaller, non-significant increase (+0.80 p.p., 95% CI: -1.09–1.08, $p = .136$). Conversely, swapping country names and GDP values produces negligible effects (all |ΔMAE| < 1.00 p.p.; all $p > .05$), indicating that models integrate structural features rather than retrieve country-specific patterns. This result reveals patterns consistent with compositional reasoning: The best-performing LLMs—Claude and Llama—combine features and flexibly integrate contradictory inputs, rather than relying on rigid memorization.

**Do findings hold across prompt formats and climate policies?**

To assess robustness across prompt formats and climate policies, we repeated the full specification using structured prompts (bullet points rather than sentences) with identical informational content. Supplementary Fig. S2 shows that results replicate across formats, though structured prompts yield systematically higher MAEs. Consistent with the main analyses, removing or flipping personal willingness produces the largest increases in MAE for Claude and Gemini, while swapping country names and GDP values induces minimal changes. Bootstrap and cross-validation diagnostics (Supplementary Fig. S3) and ablation studies (Supplementary Fig. S4) confirm these patterns are not driven by sampling variance or partition effects.

Second, bootstrap and cross-validation diagnostics show tight 95% CIs and consistent performance across resamples and data partitions (Supplementary Figs. S3–S4), suggesting that the results are not driven by sampling variance, prompt placement, or partition effects.

Third, we restrict the sample to the 112 countries with complete data for all variables, as preregistered, excluding 13 countries in which missing values were explicitly noted in the prompts. Results remain qualitatively unchanged (Supplementary Fig. S5), indicating that handling missingness does not bias predictions.

Finally, we test generalizability across 24 perception gaps about public support for a wide range of climate policies (e.g., carbon tax, renewable energy generation, Green New Deal), using three nationally representative U.S. datasets[21,23,64] (Fig. S6). Consistent with our main analysis, Claude (average MAE = 8.30 p.p., 95% CI: 6.25−10.53) and Llama (average MAE = 3.71, 95% CI: 2.73–4.78) again most accurately predict public perceptions of climate policy support; GPT performs worse than Claude and Llama (average MAE = 17.02 p.p., 95% CI: 14.25−19.88); and Gemini fails to produce any predictions, likely due to policy-safety filtering rather than predictive incapacity.

**Discussion**

Widespread perception gaps can hinder climate conversations and action[35–37]. Communicating actual levels of support can close these gaps, but efforts are most effective when targeting large misperceptions[42–44]. LLMs may offer a fast, cost-effective way to identify these gaps worldwide, helping governments and NGOs prioritize where communication efforts are urgently needed and most effective.

Across 125 countries in the Global Climate Change Survey[3], we find LLMs—particularly Claude and Llama—predict public perceptions of others' willingness to contribute financially to address climate change with high accuracy (Claude: MAE ≈ 5 p.p., $r$ = .77: Llama: ≈ 5–7 p.p., $r$ = .67). Importantly, LLMs' performance depends on whether actual public willingness (first-order willingness) based on survey data is provided: LLMs infer public perception of others' willingness primarily by treating first-order willingness as a separable, high-weight input that generalizes across countries, rather than by memorizing country-specific patterns. This aligns with the core theoretical mechanism underlying perception gaps—social projection with systematic downward bias. Nevertheless, even without providing public opinion data, LLMs retain meaningful predictive capacity, with Claude and Llama achieving mean absolute errors of approximately 7–8 p.p., using only the country's name as input.

*When are LLM predictions most reliable?* Prediction accuracy varies systematically across countries' information environments, with the highest accuracy in digitally connected, wealthier countries. This does not imply that LLMs are only useful in such settings. Rather than replacing surveys, LLMs can play complementary roles: enabling cost-effective tracking where information is abundant and providing provisional signals where resources are scarce. Building on evidence that LLMs predict personal climate beliefs[53,54], LLMs could screen both components of perception gaps—personal and perceived attitudes—across countries, flagging contexts where gaps are likely to be large. Survey resources can then be allocated dynamically: deployed intensively where predicted gaps or information scarcity (e.g., low internet penetration, limited digital media, underrepresented languages) imply high uncertainty, and used for periodic validation in information-rich settings where LLM estimates are more reliable.

*Limitations and caution.* Several limitations warrant careful interpretation. Although ablation and counterfactual tests show that LLMs rely on personal willingness rather than country names or GDP, we cannot fully rule out retrieval from training data, especially since GPT, Claude, and Llama were trained after the Global Climate Change Survey was released in April 2024. However, the observed patterns are inconsistent with pure memorization: if recall dominated, feature ablations would matter little, prediction errors would be uniform across countries, and Gemini—the most recently trained model—would perform best. Instead, removing personal willingness increases mean absolute error by 1–5 p.p., errors vary systematically with countries' information environments, and Gemini performs worst. Moreover, LLMs are trained primarily on text, whereas national perception gaps are reported only in a figure and accessed dynamically via an [interactive website](), making exposure unlikely (Supplementary S7). Overall, the results support conditional inference rather than memorization.

Second, our country-level analysis does not clarify whether these relationships hold at smaller scales—such as regions, cities, or demographic groups. Prior work suggests that perception gaps about climate policy support vary across regions in the United States, from around 10% to over 30%[21]. National estimates may hence mask intervention-relevant variation within larger countries.

*Future directions.* Several extensions could improve the reliability and scope of LLM-based perception mapping. Fine-tuning with local survey data, incorporating lightweight digital signals (e.g., social media sentiment, Google Trends), or integrating LLM inference with existing survey infrastructures (e.g., the Gallup World Poll or regional barometers) could enhance performance in low-visibility settings while enabling subnational estimation and continuous calibration. More granular validation across regions, demographic groups, and issue domains is needed to determine when LLM-generated estimates of climate perception gaps are reliable enough for resource allocation, and when direct measurement remains necessary.

Beyond technical refinement, the policy value of LLM-based perception mapping depends on its integration with intervention design. While we focus on identifying misperceptions, future work could test whether LLM-generated estimates improve intervention targeting, message framing, or timing relative to expert judgment or simple heuristics, particularly in resource-constrained settings with limited measurement capacity but large perception gaps.

## Methods

This study examines whether LLMs can predict public perceptions of others' climate action, and whether observed performance reflects structured inference from contextual information rather than direct recall of survey-specific quantities. We construct country-level measures from the Global Climate Change Survey[3], elicit predictions from multiple LLMs under controlled prompting conditions, benchmark performance against statistical models, and apply a set of mechanistic and falsification diagnostics.

The Methods follow the computational pipeline implemented in the analysis notebooks, covering data construction, prompting and inference, evaluation, diagnostic tests, and robustness analyses.

**Data sources and outcome construction**

Ground-truth data come from the Global Climate Change Survey[3] administered as part of the Gallup World Poll 2021 to 2022, covering 125 countries. The survey measured respondents' own willingness to contribute at least 1% of household income per month to address climate change and their perceptions of how willing others in their country would be to do so.

All analyses are conducted at the country level. Individual responses are aggregated using Gallup sampling weights. Pluralistic ignorance (or perception gaps) is defined as the difference between the country-level average first-order willingness and the corresponding average perceived willingness of others.

For a small number of countries, some auxiliary country-level covariates were unavailable. When such information was missing, the corresponding value was explicitly indicated as unavailable in the prompts provided to the language models. Missing covariates did not affect country inclusion, and all analyses and model evaluations are based on the full sample of 125 countries.

**Country-level inputs**

Model inputs consist of country-level demographic, economic, and climate-related factors. Sociodemographic variables, including mean age, education, and religiosity, are aggregated directly from the Gallup World Poll 2021/22 using the same sampling weights as the outcomes.

Macroeconomic indicators (GDP[65], inequality[66], HDI[67]) are drawn from the World Bank, and long-run average temperature from 2010–2019 (in degrees Celsius) from the World Bank Group's Climate Change Knowledge Portal (https://climateknowledgeportal.worldbank.org/download-data) and the CRU TS v.4.05 data (https://crudata.uea.ac.uk/cru/data/hrg/), as reported by Andre and colleagues[3]. Internet penetration is obtained from World Bank Open Data (World Bank, 2024), measured as the percentage of individuals using the internet. All variables are harmonised to the closest available year and merged by country identifier.

None of the inputs contains information on pluralistic ignorance or second-order beliefs.

**Prompting and inference protocol**

**System prompt**

All models were queried using a fixed system-level instruction that constrained responses to general reasoning based on the information provided in the prompt and explicitly discouraged reliance on memorized statistics, survey results, or academic sources. The system prompt was identical across models, countries, and prompt stages. The full verbatim system prompt is provided in the Supplementary Materials (S6).

**User prompt and staged design**

Predictions were elicited using a structured user prompt developed within a staged information framework. Across stages, the amount of information provided to the model varied from minimal inputs (e.g., country name only; stage 1) to a final full-specification condition. In the full-specification condition (stage 8), the prompt instructed models to estimate second-order beliefs, defined as respondents' beliefs about others' willingness to contribute financially to address climate change, conditional on a clearly specified information set. This set included country name, aggregated sociodemographic characteristics, macroeconomic indicators, long-run average temperature, and the country-level distribution of first-order willingness.

An example of the full-specification prompt for Argentina:

> "In Argentina, the average age of respondents is 42.1 years, 10.8% of the people have completed a tertiary education, and 52.3% say religion is important in daily life. GDP per capita (PPP, 2021) is $29,484, the top 1% holds 13.0% of total income, and 24.9% of total wealth. The Human Development Index (2021) is 0.842. The average temperature from 2010 to 2019 is 15.3°C. In this survey, 62.2% of people said they are personally willing to contribute 1% of their income each month, and an additional 14.3% would contribute a smaller amount.
>
> In a nationally representative survey with a probability-based sample of approximately 1,000 residents aged 15 and above in Argentina, respondents were asked: "Would you be willing to contribute 1% of your household income every month to fight global warming? This would mean that you would contribute 1 for every 100 of this income." Responses: Yes, No, (Don't Know), (Refused). Don't know and refused were coded as missing data. Respondents were then asked how many respondents in Argentina they think are willing to contribute at least 1% of their household income every month to fight global warming. Responses: between 0% and 100%, (Don't know), (Refused).
>
> Based on the country, socio-demographic, macro-economic indicators, temperature data, and the actual willingness data shown above, estimate what respondents in Argentina on average thought about how many **other** respondents in Argentina are willing to contribute at least 1% of their household income every month to fight global warming. Note: You are estimating people's **beliefs** about others' willingness, not the actual willingness itself. Respond with a single number between 0 and 100, with one decimal place."

Separately, as part of exploratory diagnostic analyses, we conduct ablation and perturbation tests that remove or modify individual inputs (e.g., inverting first-order willingness, permuting country names, and reassigning GDP values) to assess whether model behavior is consistent with structured inference rather than memorization. Input wording and ordering were held constant across models and countries.

**Models and decoding**

We evaluate GPT-4o mini, Claude 3.5 Haiku, Gemini 2.5 Flash, and Llama 4 Maverick (17B), using their respective APIs over the period November 2025–January 2026. All queries were executed after the public release of the Global Climate Change Survey paper (February 2024) and website (April 2024) became publicly available. According to publicly available documentation, the reported training data cutoffs are October 2023 for GPT-4o mini, July 2024 for Claude 3.5 Haiku, August 2024 for Llama 4 Maverick (17B), and January 2025 for Gemini 2.5 Flash.

Decoding was deterministic for all models, with the sampling temperature set to zero (temperature = 0), ensuring that repeated queries with identical prompts produced identical outputs. Outputs were parsed to extract numeric values, retried if invalid, and bounded to the 0 to 100 interval using a uniform procedure across models.

We acknowledge the possibility of post-publication information exposure. Rather than ruling this out a priori, we evaluate whether LLM performance is consistent with memorization of survey-specific quantities, which would imply near-exact country-level predictions, or instead reflects structured inference from the information provided, using a series of ablation, perturbation, and falsification tests described below.

**Evaluation and benchmarks**

Model performance is assessed primarily using mean absolute error (MAE) between predictions and country-level ground truth. Secondary metrics include root mean squared error (RMSE), Pearson correlation ($r$), and variance ($R^2$) explained.

For benchmarking, ordinary least squares (OLS) and Lasso regressions are estimated using the same country-level covariates provided to the LLMs. Statistical models are trained on 80 percent of countries and evaluated on a 20 percent holdout set, with repeated splits.

**Diagnostic tests**

**Sequential information addition**

In pre-registered experiments, we systematically varied the information provided to LLMs across eight stages to test how prediction accuracy evolves as features are progressively introduced. Stage 1 provides only country names as a minimal baseline. Stage 2 adds demographic variables (mean age, education, religiosity). Stage 3 incorporates macroeconomic indicators (GDP per capita, top 1% income and wealth shares, HDI). Stage 4 adds average temperature (2010-2019). Stage 5 introduces first-order climate beliefs (actual willingness to contribute). Stage 6 combines demographics and economics. Stage 7 adds temperature to the Stage 6 specification. Stage 8 provides the complete information set (all variables). This design enables assessment of how models integrate information as it accumulates and which combinations produce the most accurate predictions.

**Linear regressions of actual versus LLM-predicted outcomes**

To examine the country-level predictors of second-order beliefs (perceptions of others' willingness to contribute), we applied ordinary least squares (OLS) regressions with robust standard errors. We estimated separate models for: (1) actual second-order beliefs from survey respondents, and (2) LLM-predicted second-order beliefs from each of the four models (Claude, GPT, Llama, Gemini).

The dependent variable in all models was the country-level mean of second-order beliefs (others' willingness to contribute to climate action, 0-100 scale). Explanatory variables included country-level means of: personal willingness to contribute financially to address climate change (first-order beliefs), age, education level, religiosity, Human Development Index (HDI, 2021), GDP per capita (2021), share of income held by the top 1% (2021), and average temperature (2010-2019).

This parallel modeling approach allowed us to assess whether LLMs capture the same empirical relationships that predict human second-order beliefs, and whether they appropriately weight different predictors relative to actual human data.

**Feature ablations and counterfactuals**

To understand how models construct predictions—which information they prioritize, whether they reason structurally about social dynamics, and whether they rely on memorized data—we implement controlled feature-removal ablations and counterfactual perturbations around the full-specification (stage 8) prompt. Using the same preregistered prompts, feature-removal ablations delete one information block at a time while holding all remaining prompt content fixed. Specifically, we remove all macro-economic indicators—GDP per capita (PPP), the top 1% income share, the top 1% wealth share, and the Human Development Index (no_econ); remove religion only (no_religion); remove demographic variables while retaining GDP per capita as a coarse development proxy (no_demo); remove temperature information (no_climate); or remove all first-order willingness information (no_own_willingness). We also include a minimal country-name-only condition as a lower-bound benchmark.

We further run three counterfactual perturbations. First, we flip personal willingness by replacing the willingness block with 75% (plus 15% additional) for baseline low-willingness countries (≤ 50% willing) and with 15% (plus 10% additional) for baseline high-willingness countries (> 50% willing), holding all other inputs fixed (cf_willingness_flip). Second, we flip GDP per capita by replacing GDP per capita (PPP, 2021) with $65,000 for countries with baseline GDP per capita ≤ $20,000, and with $2,000 for countries above that threshold, leaving the remaining macro-economic indicators unchanged (cf_gdp_flip). Third, we create a name–data mismatch by pairing a rich country's covariates with a poor-country name (selected from countries with GDP per capita < $5,000) and, conversely, pairing a poor country's covariates with a rich-country name (selected from countries with GDP per capita > $40,000), while updating the question text to reference the counterfactual name (cf_name_mismatch).

Note that even with the same preregistered prompts and temperature fixed at zero, LLM predictions for nominally identical conditions—such as stage 1 of the sequential information addition and the country-name-only ablation—are not always numerically identical across runs. This residual variation reflects implementation-level nondeterminism rather than substantive differences in inputs: modern LLM inference can involve stochastic tie-breaking among near-equivalent token probabilities, low-level numerical nondeterminism from parallelized GPU computations, and periodic backend or model-snapshot updates by providers. Importantly, this variation is small relative to the effects induced by feature removal or counterfactual perturbations and does not affect the qualitative patterns or conclusions drawn from the ablation and counterfactual analyses.

**Structural heterogeneity**

Each LLM's prediction error (MAE) is regressed on internet penetration and GDP per capita using ordinary least squares (OLS) without covariates.

**Robustness and replicability**

Robustness checks also include repeating the analysis across an extended set of 24 alternative perception gaps drawn from three nationally representative U.S. datasets[21,23,64]. For these analyses, LLMs were queried using the same prompts as in our original tests, adapted to the specific policy. The 24 alternative questions are provided in the Supplementary Materials (S8).

**Code and preregistration**

Gallup World Poll microdata are subject to access restrictions, but all aggregation and analysis scripts are provided. The study was pre-registered at https://osf.io/ez4ur/overview?view_only=e8add9f07f9148dfad359eed89744d5e.

**AI Usage**

Artificial intelligence tools, including ChatGPT (OpenAI) and Grammarly, were used to polish language, enhance clarity, and organise content in the manuscript. No AI tools were used to produce scientific findings or perform analysis. All outputs were carefully reviewed and verified by the authors.

# Supplementary Materials

## S1: System Instruction for LLM Queries

To prevent data leakage and ensure that predictions rely on inference rather than retrieval of memorized survey results, all LLM queries included the following system instruction:

"You are a prediction assistant making estimates based ONLY on the information provided in this specific prompt.

**CRITICAL INSTRUCTIONS:**

1. Do NOT cite, reference, or mention ANY research papers, academic studies, surveys, or authors (including but not limited to André et al., Sparkman et al., Leviston et al., Lees et al., or any other researchers).
2. Do NOT use any memorized data, statistics, or percentages from your training about climate change opinions, pluralistic ignorance, or survey results.
3. Treat this as a completely NOVEL scenario—ignore any similar studies you may have seen during training.
4. Do NOT reference phrases such as 'research shows,' 'studies indicate,' 'surveys have found,' or similar language.
5. Base your estimate ONLY on:
    - General reasoning about human psychology and behavior
    - The specific information provided in this prompt
    - First principles about how people form beliefs about others

*Your task is to predict what percentage people THINK others believe (second-order belief), not what people actually believe (first-order belief). This is a prediction task requiring general reasoning, not recall of specific research findings.*

**Response format:** Provide ONLY a JSON object containing a single number between 0 and 100 with one decimal place: `{"prediction": XX.X}`

Do not include any explanation, reasoning, or text—only the JSON.

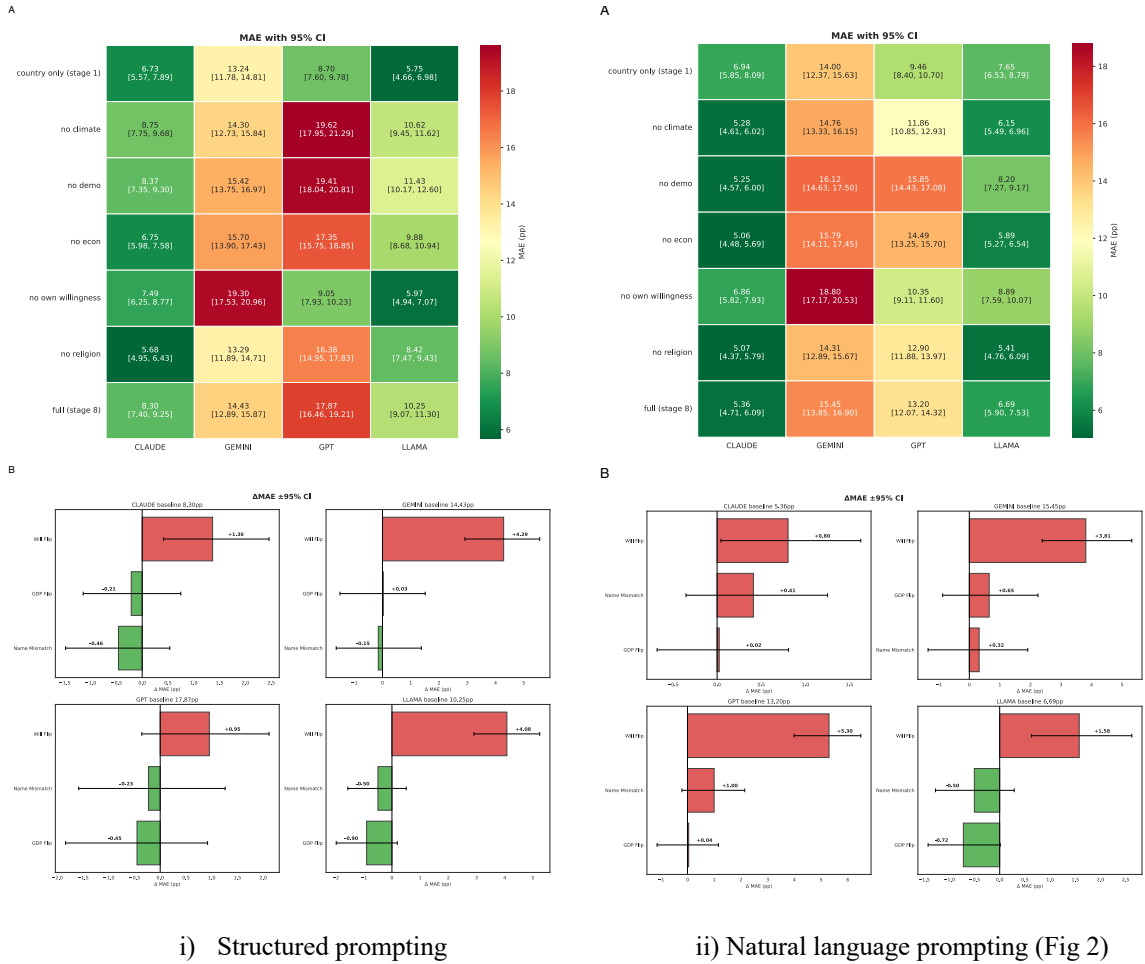

i) Structured prompting ii) Natural language prompting (Fig 2)

**Fig. S2: Robustness to prompt format: Structured vs. natural language presentation.** Comparison of LLM performance using structured (bullet-point) versus natural-language prompts. Panels show mean absolute error (MAE) for ablation tests and counterfactual manipulations under both prompt formats. Country attributes (demographics, GDP, climate indicators, first-order beliefs) were identical across formats; only presentation style varied. Results demonstrate that core findings—particularly the importance of first-order willingness and robustness to GDP/name swaps—replicate across prompt structures, indicating that the behavioral mechanism does not depend on narrative presentation.

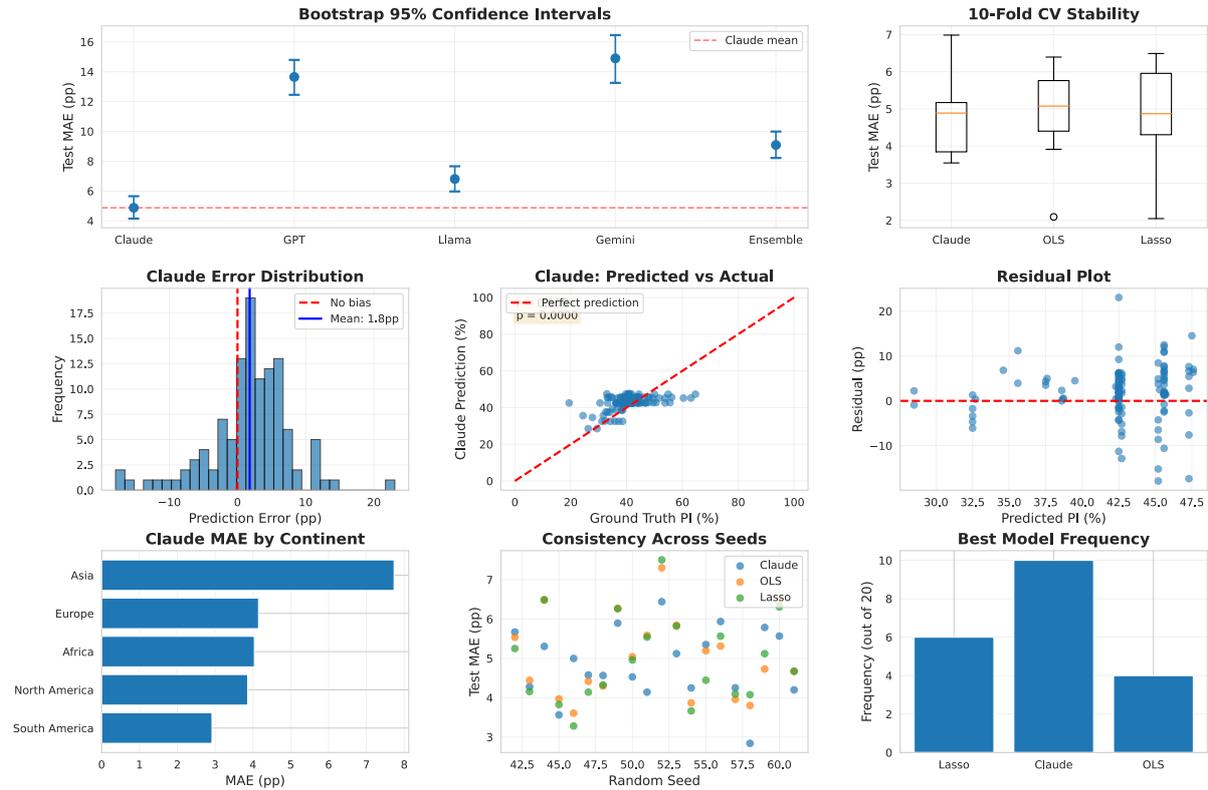

**Fig. S3: Bootstrap and cross-validation diagnostics for Claude.** Robustness diagnostics for Claude's predictions across 125 countries. Top left: Bootstrap 95% confidence intervals for mean absolute error (MAE) across 10,000 resamples, comparing Claude with other models (GPT, Llama, Gemini, Ensemble). Tight intervals indicate results are not driven by sampling variance. Top right: 10-fold cross-validation stability showing consistent performance across data partitions. Middle left: Distribution of Claude's prediction errors (mean = 1.8 p.p. bias), demonstrating approximately normal error distribution. Middle center: Predicted vs. actual values showing a strong linear relationship ($r = 0.77$, $p < .001$) with minimal systematic bias. Middle right: Residual plot confirming homoscedasticity (constant variance) across prediction range. Bottom left: MAE by continent, showing systematic variation with information environments (lowest in Europe, highest in Asia). Bottom center: Consistency across random seeds, demonstrating stable performance. Bottom right: Best model frequency across bootstrap iterations, confirming Claude as the most consistently accurate model.

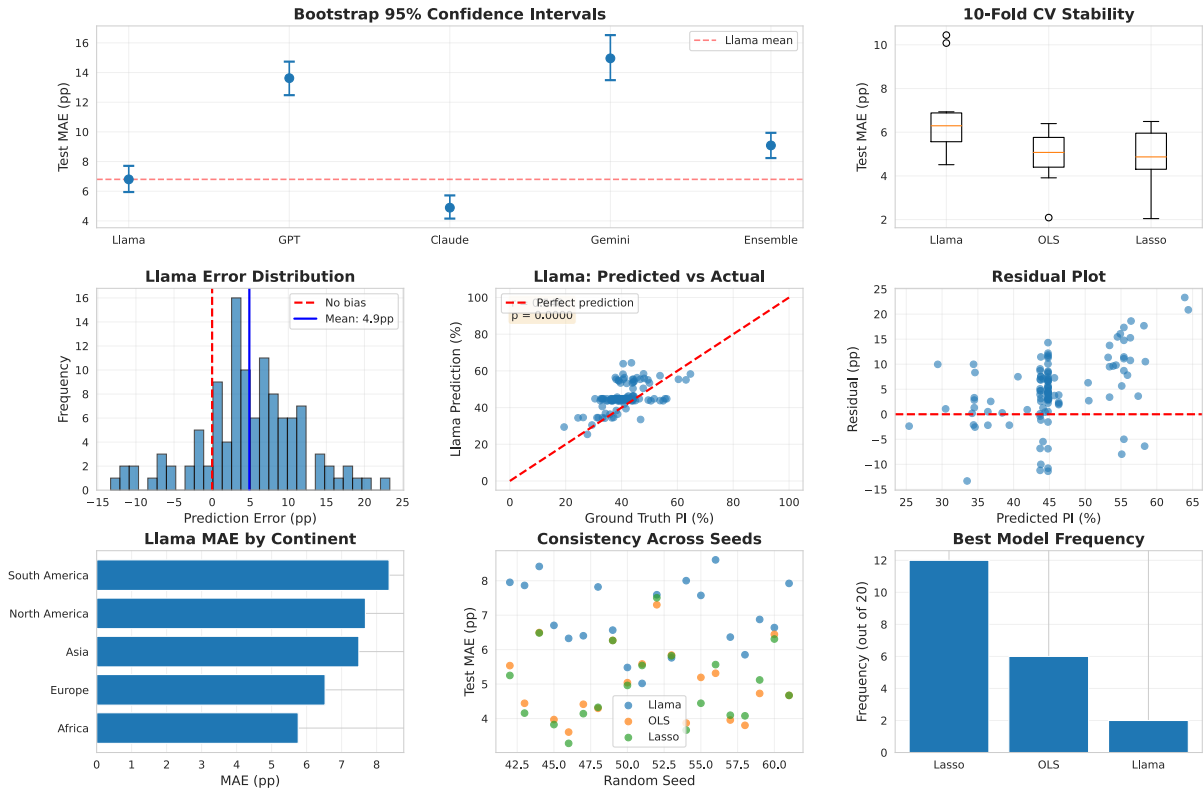

**Fig. S4: Bootstrap and cross-validation diagnostics for Llama.** Robustness diagnostics for Llama's predictions across 125 countries. Top left: Bootstrap 95% confidence intervals for mean absolute error (MAE) across 10,000 resamples, comparing Llama with other models (Claude, GPT, Gemini, Ensemble). Tight intervals indicate results are not driven by sampling variance. Top right: 10-fold cross-validation stability showing consistent performance across data partitions. Middle left: Distribution of Llama's prediction errors (mean = 4.9 p.p. bias), demonstrating approximately normal error distribution. Middle center: Predicted vs. actual values showing a strong linear relationship ($r = 0.67$, $p < .001$) with minimal systematic bias. Middle right: Residual plot confirming homoscedasticity (constant variance) across prediction range. Bottom left: MAE by continent, showing systematic variation with information environments (lowest in Europe and Africa, highest in South America). Bottom center: Consistency across random seeds, demonstrating stable performance. Bottom right: Best model frequency across bootstrap iterations, confirming Llama as the second most consistently accurate model after Claude.

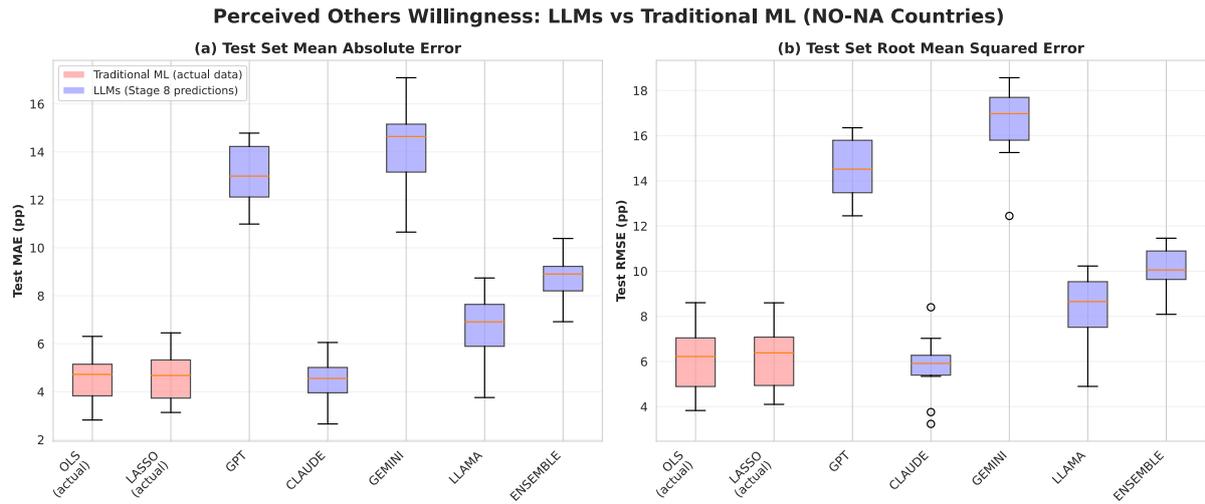

**Fig. S5: LLM Performance vs. traditional machine learning (ML) on complete-data sample.** Comparison of LLM predictions (stage 8) against ordinary least squares (OLS) and LASSO regression trained on actual country-level data, restricted to 112 countries with complete information for all variables (excluding 13 countries with any missing values). Models were evaluated using 10 repeated 80:20 train-test splits stratified by continent. (a) Test set mean absolute error (MAE). (b) Test set root mean squared error (RMSE). Boxplots show distribution across 10 iterations. Traditional ML models (red) are trained on actual data; LLMs (blue) rely on prompted inference without training.

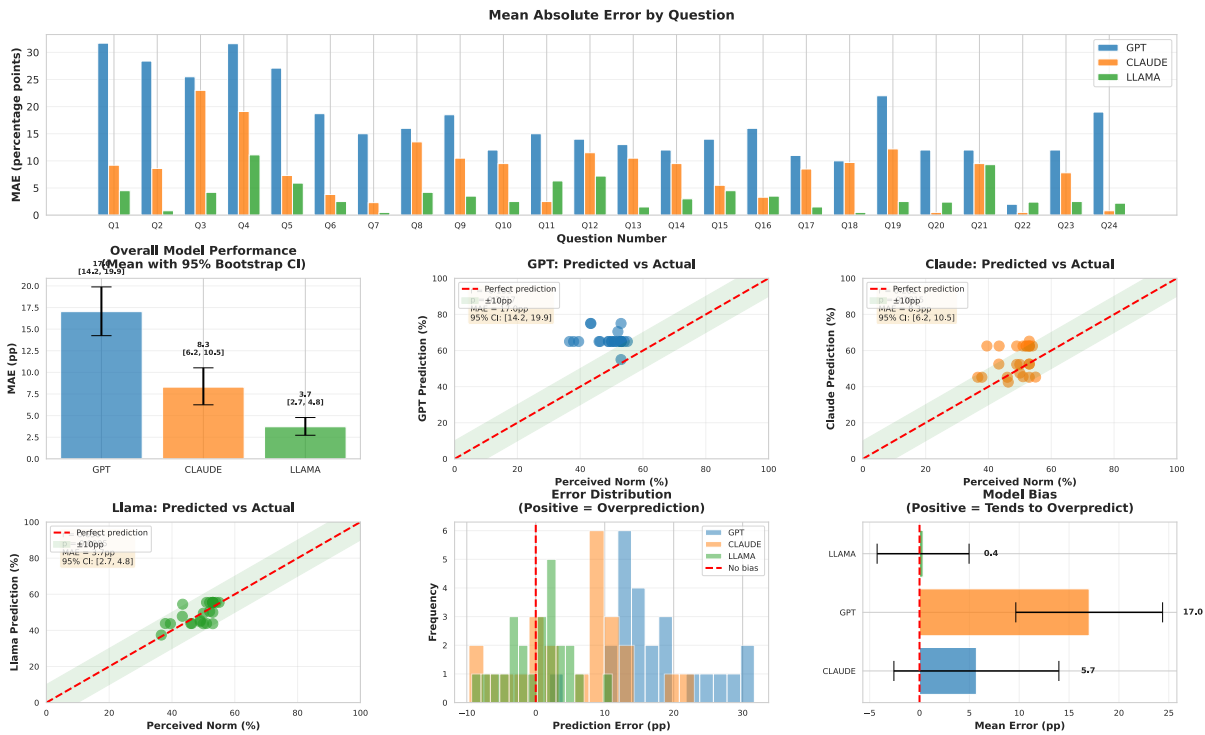

**Fig. S6: Generalization to U.S. climate policy (24 items).** Model performance predicting public perceptions of others' support across 24 climate policy items from three nationally representative U.S. datasets[21,23,64]. Top panel: Mean absolute error (MAE) by question for GPT, Claude, and Llama across all 24 items. Middle left: Overall model performance (mean with 95% bootstrap CI). Middle center and right: Predicted vs. actual scatter plots for GPT and Claude showing correlations with ground truth. Bottom left: Predicted vs. actual for Llama. Bottom center: Error distribution across all predictions, showing frequency of overestimation vs. underestimation. Bottom right: Model bias (mean error), with positive values indicating overestimation and negative values indicating underestimation. Gemini produced no predictions for any items, likely due to content filtering of policy-related queries.

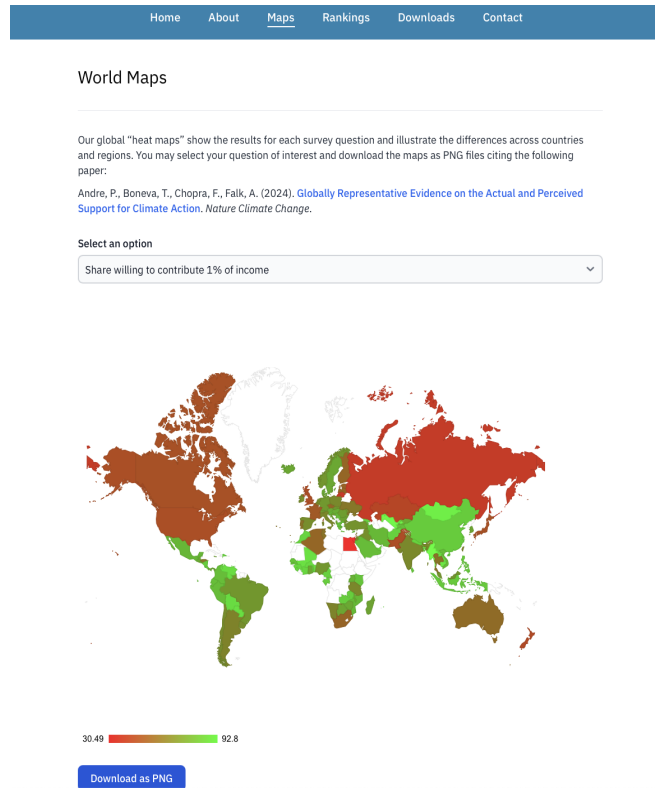

**a) Map page**

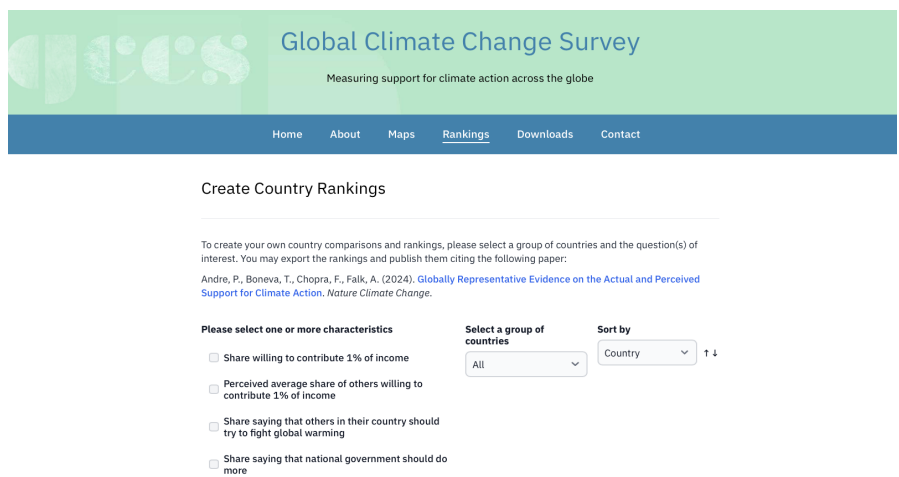

**b) Rankings page**

**Fig. S7: GCCS website interactive data access interfaces.** (a) Maps page (https://gccs-survey.org/maps): Interactive heat map showing second-order beliefs about willingness to contribute 1% of income toward climate action. Country-level values are revealed only through individual hover interactions. The color gradient (30.49-92.8% range shown) provides relative comparisons but not precise numerical values. (b) Rankings page (https://gccs-survey.org/rankings): Custom ranking generator requiring users to select survey questions, country groups, and sorting criteria before generating exportable country comparisons. No pre-generated rankings or comprehensive datasets are available without active user configuration. Both interfaces require interactive engagement to access country-specific data, limiting the availability of systematic country-level rankings during potential model training.

**S8: Issue domains for U.S. climate policy generalization tests**

To assess whether models' ability to predict public perceptions of others' climate action generalizes beyond the Gallup World Poll measure, we tested predictions across 24 climate policy belief items from three U.S. datasets: Sparkman et al. (2022), Lees et al. (2023), and Swim & Baker (2025). All prompts followed identical structure: (1) survey context and sample description, (2) first-order belief question with response options, (3) second-order belief question asking what respondents think others believe, (4) explicit instruction to estimate beliefs about others' beliefs rather than personal views, and (5) request for single numerical response between 0-100 with one decimal place.

| PI | Topic & Question Type |
|---|---|
| PI1 | Worry about climate change (≥ somewhat worried) |
| PI2 | Carbon tax to reduce other taxes (national sample) |
| PI3 | Carbon tax (variant labelled "100% RE") |
| PI4 | Renewable energy on public lands (national sample) |
| PI5 | Green New Deal |
| PI6 | Climate change is a *very big* problem |
| PI7 | Regulate $CO_2$ as a pollutant |
| PI8 | Require 100% renewable electricity by 2035 |
| PI9 | Fund more research into renewable energy |
| PI10 | Tax rebates for EVs and solar |
| PI11 | Carbon tax (registered voters version) |
| PI12 | Transition the U.S. economy to 100% clean energy by 2050 |
| PI13 | Tax incentives for clean appliances |
| PI14 | Tax incentives for building efficiency retrofits |
| PI15 | Federal funding for energy efficiency in low-income housing |
| PI16 | Funding for low-income & minority communities harmed by pollution |
| PI17 | Opposition to expanding offshore drilling |
| PI18 | Opposition to fossil fuel extraction on public lands |
| PI19 | Renewable energy on public lands (registered voters) |
| PI20 | Re-establish the Civilian Conservation Corps |
| PI21 | Jobs program: close abandoned oil & gas wells |
| PI22 | Jobs program: rehabilitate old coal mines |
| PI23 | Funding to help farmers store more carbon in soil |
| PI24 | President declaring climate emergency (if Congress doesn't act) |